**Magnetotransport in $Fe_3O_4$ nanoparticle arrays dominated by non-collinear surface spins**

Seongjin Jang, Wenjie Kong, Hao Zeng[1]

Department of Physics

University at Buffalo, the State University of New York, Buffalo, NY 14260

Magnetotransport in arrays of monodisperse magnetite nanoparticles has been studied as a function of annealing temperatures. Charge transport mechanisms change from thermally assisted interparticle tunneling to hopping between Fe-sites within the particle as the interparticle spacing is decreased. Despite this difference, magnetoresistance (MR) as a function of field shows a ubiquitous behavior dominated by non-collinear surface spins. All MR as a function of field can be fitted accurately by a Langevin-like function.

[1]Electronic mail: haozeng@buffalo.edu

## I. INTRODUCTION

Charge transport in magnetic granular systems consisting of magnetic grains embedded in a nonmagnetic matrix is often spin dependent. They exhibit either giant magnetoresistance (GMR) for a conducting matrix or tunneling magnetoresistance (TMR) for a dielectric matrix.[1-3] The underlying physics of granular TMR was proposed to be the exchange splitting of electronic bands of the ferromagnetic grains, leading to difference in tunneling probability of electrons with different spin orientations.[4] The conductance was given by Inoue *et al*. as $G \propto G_0(1+P^2\cos\theta)\exp(-2\kappa s)\exp(-E_C/k_BT)$, where $P$ is the spin polarization, $\theta$ the angle between the magnetization of two grains, $\kappa$ the wave vector within the barrier, $s$ the barrier width, and $E_C$ the charging energy of the grain.[5] The average $\langle\cos\theta\rangle$ in the limit of uncorrelated magnetization of the grains is $(M/M_S)^2$. In general, the MR scales with the global magnetization in the form of $F[(M/M_S)]$, where $F$ is an even function.[6] Although the origin of granular GMR is spin dependent scattering, its dependence on magnetization orientation of grains and hence the scaling of MR with magnetization remain unchanged.

Magnetotransport in granular systems has been extensively studied.[1, 3, 7-13] Most of the measurements are done on sputtered composite films or compacted/sintered powders. Due to the large distribution in grain size and spacing, it is generally difficult to extract properties related to dimensional parameters. Despite the large difference in material systems, the magnetotransport can show some commonalities. For example, the MR often exhibits a high field component, which does not appear to scale with the magnetization.[7, 8, 10, 11, 13] It has been attributed to interface or surface scattering/tunneling in some early work,[7, 11] but a systematic analysis of the surface/interface effects is lacking.

In this paper we show magnetotransport behavior of a self-assembled 3D array of $Fe_3O_4$ nanoparticles, with varying annealing temperatures ($T_a$). By controlled annealing, the interparticle distance can be systematically varied, so that changing of charge transport mechanisms from thermally assisted interparticle tunneling to intra particle hopping between $Fe^{2+}$ and $Fe^{3+}$ sites is observed. Correspondingly the MR behavior changes from tunneling controlled to scattering controlled. Despite such differences, however, MR of all samples shows strong field dependence that can be universally fitted by a Langevin-like function. We propose that this field dependence originates from spin dependent scattering/tunneling at nanoparticle surfaces and interfaces, where spin canting occurs.

## II. EXPERIMENT

Monodisperse $Fe_3O_4$ nanoparticles with 6 nm diameter were synthesized by high temperature solution phase reaction, as reported in earlier work.[14] A small amount of the nanoparticle solution was deposited on a Si substrate with a 1 μm thermally oxidized $SiO_2$. Controlled evaporation of the solvent led to a self-assembled nanoparticle array with interparticle distance of about 3-4 nm determined by the chain length of the surfactant molecules. The TEM image of an ordered array of as-deposited $Fe_3O_4$ nanoparticles on TEM grids is shown in fig. 1(a). The nanoparticle arrays were then annealed under nitrogen for 1 hr, with temperatures ranging from 100 °C to 650 °C. The annealing removed surfactants and reduced interparticle spacing, with the ordering of the assembly well preserved up to 500 °C (inset of fig. 1(b). This was done on a particle array deposited on a $SiO_2$ substrate, with the deposition condition identical to that used for transport measurements). At temperatures above 600 °C, particles started to neck together and neighboring particles would share a common interface, as can be seen from fig. 1(b). No significant grain growth was observed even at 650 °C. For charge transport studies, a pair of lateral gold electrodes was fabricated by photolithography with gap spacing from 2 to 15 μm. Nanoparticles were then self-assembled between the electrodes and annealed.

## III. RESULTS AND DISCUSSION

### A. Magnetic properties

As-deposited samples are superparamagnetic with a blocking temperature ($T_B$) of about 40 K. As $T_a$ increases, $T_B$ increases accordingly due to the reduced interparticle spacing and enhanced magnetostatic interactions, as can be seen from the shift in the peak temperatures of Zero-field-cooled magnetization curves in fig. 2.[15] Samples annealed at above 400 °C are ferromagnetic ($T_B$ > 300 K). For samples with high $T_B$, a sharp drop in magnetization at around 120 K is observed as the samples being cooled. This temperature matches well with the Verwey transition temperature ($T_V$) of bulk magnetite of 125 K,[16] suggesting that our nanoparticles contain nearly stoichiometric $Fe_3O_4$. We notice that there is a broad peak-like feature for samples annealed at and above 600 °C. We are not clear of its origin but speculate that a portion of the particles have lower $T_B$. Magnetic hysteresis measurements suggest that as $T_a$ is above 400 °C, samples exhibit ferromagnetic behavior. All samples show a small susceptibility at fields higher than 1 Tesla. This behavior could be related to non-collinear spins at the particle surface, as will be discussed below.

### B. Temperature dependent conductance

Un-annealed samples are insulating due to the large interparticle spacing with dielectric surfactants in between. Samples annealed above 200 °C show thermally activated conduction characterized by the exponential temperature dependence of conductivity. The zero-bias-voltage conductance ($G_0$) as a function of temperature obtained from I-V measurements are shown in fig. 3. For samples annealed below 500 °C, $G_0$ vs. T can be well fitted by $C\exp[(E_C/kT)^{1/2}]$, as shown in Fig. 3(b). This is consistent with the physical picture of thermally assisted interparticle tunneling,[17] where the charging energy of a single nanoparticle dominates the temperature dependence of the tunnel conductance. The power index 1/2 indicates a distribution of the energy barrier, resulting e.g. from the variation in interparticle spacing.

For $T_a$ above 500 °C, a kink appears at around $T_V$ of about 120 K as seen from the $G_0$-T data (fig. 3(a)), separating two regimes with different temperature dependence. $G_0$ above $T_V$ shows an $\exp(E_B/T)$ dependence (fig.3(c)). Below $T_V$, $G_0$-T relation is best fitted by $\exp[(T_0/T)^{1/4}]$ (fig. 3(d)). It is theoretically proposed that Verwey transition is accompanied by a change from long range charge ordering below $T_V$ to short range ordering above $T_V$, and charge transport occurs through thermally activated hopping of small polarons.[16] The transition can then be best described as a semiconductor-semiconductor transition with two states of different conductivity. Above $T_V$, because of the short range charge ordering, charges always find highest probability of hopping between nearest neighbor $Fe^{2+}$ and $Fe^{3+}$ sites. The $T^{-1}$ dependence of conductance is consistent with this nearest neighbor hopping mechanism. The energy barrier $E_B$ obtained from the fitting of the $T_a$ = 650 °C curve (fig. 3(b)) is 40meV, well within the range of reported values of 20-60 meV.[18] Below $T_V$, long range charge order leads to much higher hopping barrier, and charges can find more conducting paths through non-nearest neighbors. The $T^{-1/4}$ dependence of conductance is consistent with this variable range hopping behavior.[19] From the fitting $T_0$ is determined to be $1.0\times10^8$ K, which is on the same order of magnitude but smaller than previously reported values for $Fe_3O_4$ thin films.[20, 21]

The fact that samples annealed with different temperatures show different transport behavior mainly originates from difference in interparticle spacing. For low temperature annealed samples, transport is dominated by interparticle tunneling. The hopping energy barrier between different Fe-sites within the particle is relatively small compared to the interparticle tunneling barrier. Therefore resistance due to intra particle hopping is of secondary importance, and no Verwey transition is observed in samples with low annealing temperatures.

**C. Magnetotransport-temperature dependence**

Temperature dependence of MR defined as (R(H)-R(H=0))/R(H=0) for samples annealed at different temperatures also shows different behavior, which is clearly seen in fig. 4: for samples annealed at low temperatures showing thermally assisted tunneling, the magnitude of MR increases monotonically with decreasing temperature (fig. 4(a)). For samples annealed at above 500 °C showing the Verwey transition, the magnitude of MR exhibits a maximum at $T_V$, and decreases with both increasing and decreasing temperatures (fig. 4(b)).

However, as spin dependent resistance $\Delta R=|R_H-R_0|$ is plotted, it increases monotonically with decreasing temperature, regardless of the annealing temperature. As seen from fig. 4(c) and (d), ΔR vs. T shows approximately exponential temperature dependence for both $T_a$ = 400 and 650 °C. This is different from metallic granular systems where ΔR is approximately linear in temperature.[1] The exponential behavior suggests a thermal activation process, maybe related to thermally assisted spin flipping. When considering MR, the value is normalized by total resistance which includes contributions from spin independent part. Because the total resistance increases with decreasing temperature faster than the increase of ΔR for $T<T_V$, it leads to an upturn in MR vs. T for high temperature annealed samples.

**D. Magnetotransport-field dependence**

Although resistance and MR show different temperature dependence for samples annealed at different temperatures, the MR as a function of field exhibits a ubiquitous behavior. As seen from fig. 5(a) and (b), the magnitude of MR increases with increasing applied field, changing steeper at lower fields yet showing no sign of saturation even at the maximum applied field of 5 Tesla. Such field dependence is observed for all samples annealed at different temperatures, regardless of the charge transport mechanisms, and independent of the magnitude of MR values. It is also qualitatively similar to MR behavior reported earlier in granular films.[7, 8, 11, 21, 22] More importantly, comparing our magnetic hysteresis with MR vs. field, we clearly see that MR is not correlated with global magnetization. As a comparison, from 1 T to 5 T, the change in magnetization is less than 5% (inset of fig. 5(b)), while that of MR is more than half of the total value. The small susceptibility in the magnetization curve at fields greater than 1 T cannot explain the large change in MR in this field range.

Several earlier studies on granular systems attributed the high field dependence to the presence of superparamagnetic particles.[23, 24] Due to inevitable distribution in grain sizes in those granular systems, smaller grains are superparamagnetic while larger ones are ferromagnetic. Since the magnetization of superparamagnetic particles can be described by Langevin's function (assuming random easy axis orientation with low anisotropy), when such grains make main contributions to spin dependent transport, MR can be described by Langevin's function.

In our system, the particles are monodisperse with size distribution δd/d less than 10%. We do not expect a progressive unblocking of particles due to large size distributions. Furthermore, the superparamagnetic blocking temperature is above room temperature for samples annealed above 400 °C. We propose that the MR as a function of field for our system originates from a layer with non-collinear spins at the nanoparticle surface. Surface with non-collinear spins is commonly observed in nanoparticle systems with competing exchange interactions. Due to lower symmetry and missing nearest neighbors, antiferromagnetic exchange contributes stronger at the surface, leading to lower magnetic order.[25, 26] The competing interactions also lead to weaker exchange coupling between the surface and particle interior. Thus the surface magnetization can have its own excitation modes different from that of bulk, and its field dependence can be described by a Langevin's function $L(\alpha H)$. When charges traverse the nanoparticle array, they will experience two types of spin dependent scattering (or tunneling): 1. within the nanoparticle between a shell with non-collinear spins and spin-aligned core; 2. at the boundaries of two particles, between two shells with non-collinear spins. In the simplest approximation, MR is proportional to $\langle \cos\theta_{ij} \rangle$ between the two magnetic entities. For the first process, the MR value is proportional to the average magnetization of the shell with non-collinear spins which can be described by $L(\alpha_1 H)$. For the second process, the value is proportional to $L^2(\alpha_2 H)$. Based on the assumption above, MR can be expressed by the following equation:

$$MR(H) = P_1 L(\alpha_1 H) + P_2 L^2(\alpha_2 H) \qquad (1),$$

where $L(\alpha H)$ is the Langevin's function

$$L(\alpha H) = \frac{1}{\tanh(\alpha H)} - \frac{1}{\alpha H} \qquad (2)$$

and $P_1$, $P_2$, $\alpha_1$ and $\alpha_2$ are fitting parameters. $\alpha_i = N_i \mu_B / k_B T$ is related to the effective number of correlated spins $N_i$ participating in the spin dependent transport. In fig. 5(a) and (b) the solid lines are the fitting curves using Equation 1. This function fits accurately all the MR curves of those samples annealed at different temperatures, including those published in our previous work.[17] This indicates that the two proposed spin scattering processes make dominating contributions to MR of the nanoparticle arrays. The parameters $P_1$ and $P_2$ represent the weighted contribution of the two processes to MR. $P_1$ and $P_2$ as a function of temperature for $T_a$= 400 °C is shown in fig. 5(c), both of which increases with decreasing temperature. $P_1$ is always larger than $P_2$. This can be understood since for a given charge, it traverses more often the interfaces between the nanoparticle surface and core than the boundaries between two particles. $P_1$ and $P_2$ contain information regarding temperature dependent spin polarization at the nanoparticle surfaces and interfaces. The fitted $\alpha_i$ values are approximately temperature independent. For example, for $T_a$ = 400 °C, $\alpha_1 = 4.4\times10^{-5}$ Oe$^{-1}$ and $\alpha_2 = 3.5\times10^{-4}$ Oe$^{-1}$. This means that $N_i$ decreases quasi-linearly with temperature. As can be seen from fig. 5(d), $N_1$ changes from 180 at 300 K to 30 at 70 K and $N_2$ varies from 1500 to 300. This can be understood qualitatively since surface spin non-collinearity will decrease with decreasing temperature due to weakened thermal fluctuations. Efforts are underway to correlate $P_i$ and $N_i$ with microscopic spin configurations and to understand the role of interparticle exchange coupling. The spin canting combined with the lower spin polarization at nanoparticle surfaces[27] may be the main factors contributing to lower MR than that predicted for a half-metal.

## IV. CONCLUSION

In conclusion, we have studied charge transport in $Fe_3O_4$ nanoparticle arrays annealed at different temperatures. For low annealing temperature, the transport is dominated by interparticle tunneling; while for high annealing temperature, Verwey transition separates two regimes: nearest neighbor hopping for $T > T_V$ and variable range hopping for $T < T_V$. Despite such difference, magnetotransport is dominated by misaligned spins at nanoparticle surfaces and field dependence of MR can be fitted by a Langevin-like function.

## ACKNOWLEDGMENTS

We would like to thank the fruitful discussion with Renat Sabirianov. This work is supported by NSF DMR-0547036 and UB MIPS program.

## References

[1] C. L. Chien, J. Q. Xiao, and J. S. Jiang, Journal of Applied Physics **73**, 5309 (1993).

[2] E. Z. Meilikhov, Jetp Letters **69**, 623 (1999).

[3] J. F. Wang, J. Shi, D. C. Tian, H. Deng, Y. D. Li, P. Y. Song, and C. P. Chen, Applied Physics Letters **90**, 213106 (2007).

[4] E. Y. Tsymbal, O. N. Mryasov, and P. R. LeClair, Journal of Physics-Condensed Matter **15**, R109 (2003).

[5] J. Inoue and S. Maekawa, Physical Review B **53**, 11927 (1996).

[6] C. L. Chien, Annual Review of Materials Science **25**, 129 (1995).

[7] Z. L. Lu, W. Q. Zou, L. Y. Lv, X. C. Liu, S. D. Li, J. M. Zhu, F. M. Zhang, and Y. W. Du, Journal of Physical Chemistry B **110**, 23817 (2006).

[8] M. Venkatesan, S. Nawka, S. C. Pillai, and J. M. D. Coey, Journal of Applied Physics **93**, 8023 (2003).

[9] P. Poddar, T. Fried, and G. Markovich, Physical Review B **65,** 172405 (2002).

[10] L. Savini, E. Bonetti, L. Del Bianco, L. Pasquini, L. Signorini, M. Coisson, and V. Selvaggini, Journal of Magnetism and Magnetic Materials **262**, 56 (2003).

[11] C. Park, Y. G. Peng, J. G. Zhu, D. E. Laughlin, and R. M. White, Journal of Applied Physics **97,** 10C303 (2005).

[12] P. Chen, D. Y. Xing, Y. W. Du, J. M. Zhu, and D. Feng, Physical Review Letters **87,** 107202 (2001).

[13] O. Chayka, L. Kraus, P. Lobotka, V. Sechovsky, T. Kocourek, and M. Jelinek, Journal of Magnetism and Magnetic Materials **300**, 293 (2006).

[14] S. H. Sun and H. Zeng, Journal of the American Chemical Society **124**, 8204 (2002).

[15] S. Morup and E. Tronc, Physical Review Letters **72**, 3278 (1994).

[16] F. Walz, Journal of Physics-Condensed Matter **14**, R285 (2002).

[17] H. Zeng, C. T. Black, R. L. Sandstrom, P. M. Rice, C. B. Murray, and S. H. Sun, Physical Review B **73,** 020402(R) (2006).

[18] B. Raquet, J. M. D. Coey, S. Wirth, and S. von Molnar, Physical Review B **59**, 12435 (1999).

[19] P. A. Lee, Physical Review Letters **53**, 2042 (1984).

[20] S. B. Ogale, K. Ghosh, R. P. Sharma, R. L. Greene, R. Ramesh, and T. Venkatesan, Physical Review B **57**, 7823 (1998).

[21] G. Q. Gong, A. Gupta, G. Xiao, W. Qian, and V. P. Dravid, Physical Review B **56**, 5096 (1997).

[22] A. Maeda, M. Kume, S. Oikawa, and K. Kuroki, 37th Annual conference on magnetism and magnetic materials **76**, 6793 (1994).

[23] B. J. Hickey, M. A. Howson, S. O. Musa, and N. Wiser, Physical Review B **51**, 667 (1995).

[24] D. Barlett, F. Tsui, D. Glick, L. Lauhon, T. Mandrekar, C. Uher, and R. Clarke, Physical Review B **49**, 1521 (1994).

[25] J. M. D. Coey, Physical Review Letters **27**, 1140 (1971).

26 X. Batlle and A. Labarta, Journal of Physics D-Applied Physics **35**, R15 (2002).
27 R. Pentcheva, F. Wendler, H. L. Meyerheim, W. Moritz, N. Jedrecy, and M. Scheffler, Physical Review Letters **94**, 126101 (2005).

**Figure captions:**

Figure 1(a) A typical TEM image of an ordered array of as-deposited $Fe_3O_4$ nanoparticles; (b) a TEM image of a $Fe_3O_4$ nanoparticle array annealed at 650 °C; insert: a TEM image of a $Fe_3O_4$ nanoparticle array annealed at 500 °C.

Figure 2 Zero-field-cooled magnetization as a function of temperature for samples (a) without annealing and annealed at (b) 300 °C, (c) 600 °C and (d) 650 °C.

Figure 3(a) Resistance plotted in logarithmic scale as a function of temperature for $T_a$ = 400 and 650 °C; (b) Zero-bias-voltage conductance ($G_0$) plotted in logarithmic scale as a function of $T^{-1/2}$ for $T_a$ = 400 °C; (c) $G_0$ from 130 K to 300 K as a function of $T^{-1}$ for $T_a$ = 650 °C; (d) $G_0$ from 50 K to 110 K as a function of $T^{-1/4}$ for $T_a$ = 650 °C.

Figure 4 MR as a function of temperature for $T_a$ = (a) 400 °C and (b) 650 °C; $\Delta R$ in logarithmic scale as a function of temperature for $T_a$ = (c) 400 °C and (d) 650 °C.

Figure 5 MR as a function of field measured at different temperatures for $T_a$ = (a) 400 °C and (b) $T_a$ = 650 °C, inset: 300 K hysteresis loop measured at the same field range; (c) $P_1$ and $P_2$ as a function of temperature for $T_a$ = 400 °C, (d) $N_1$ and $N_2$ as a function of temperature for $T_a$ = 400 °C.

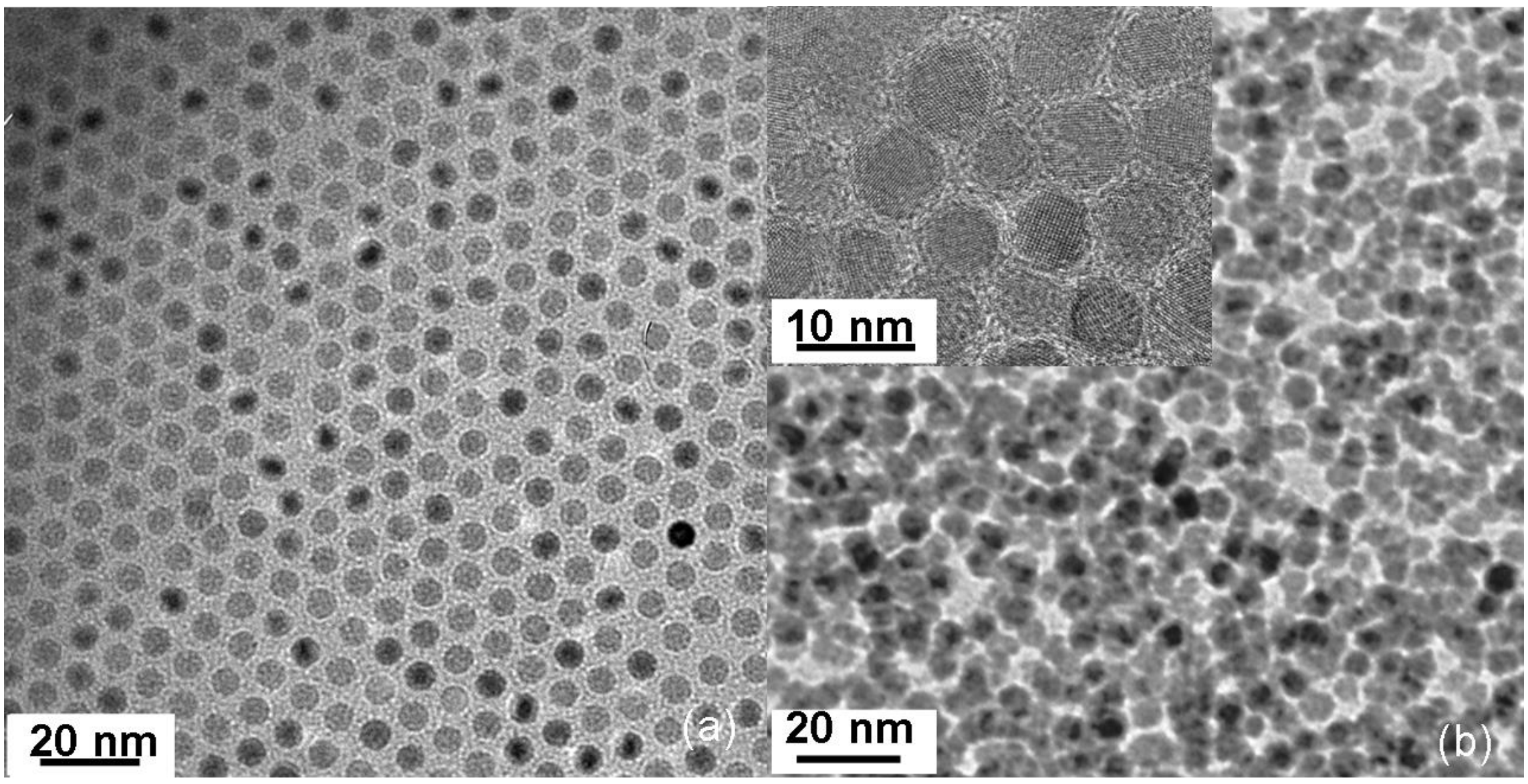

Fig. 1

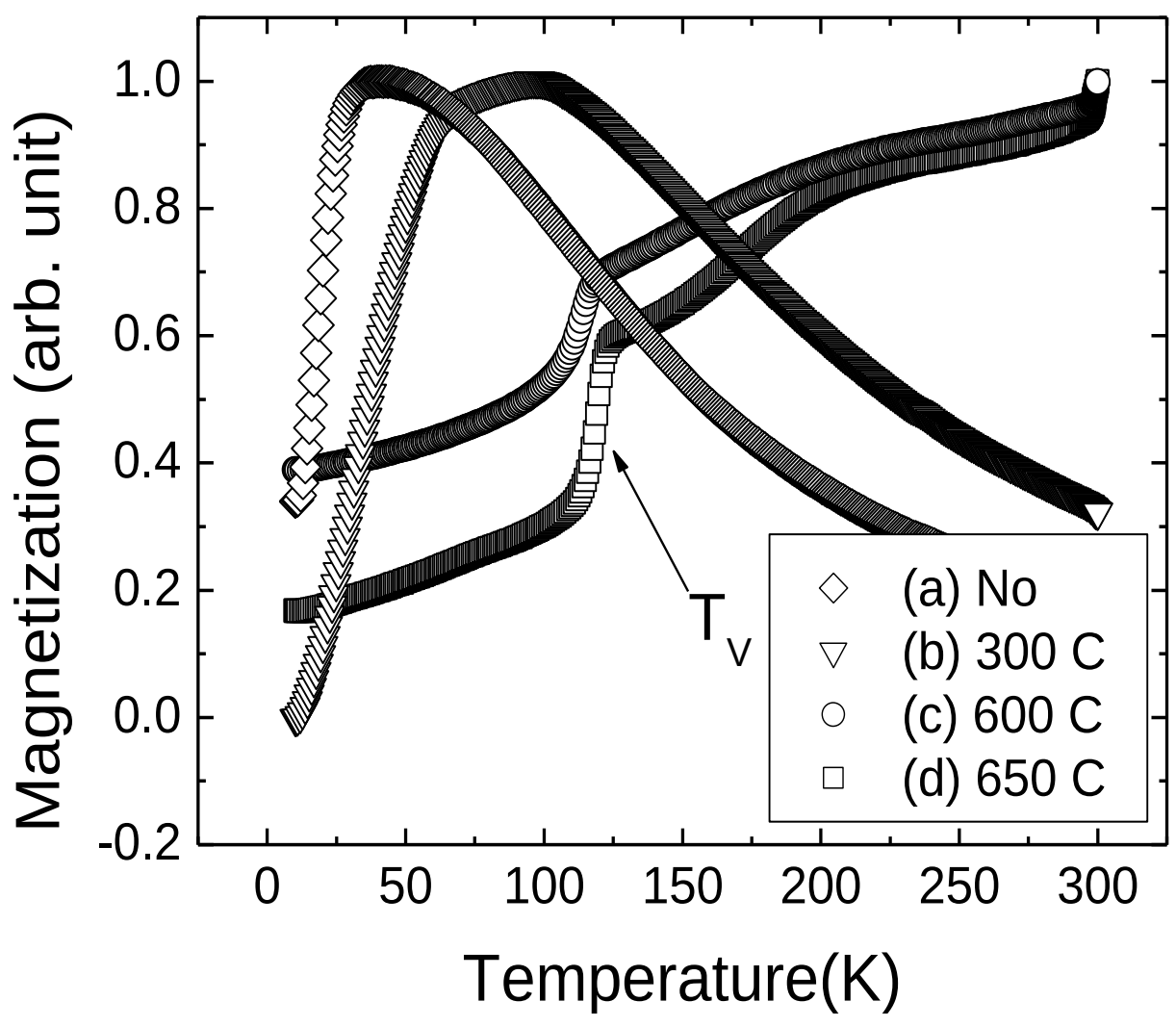

Fig. 2

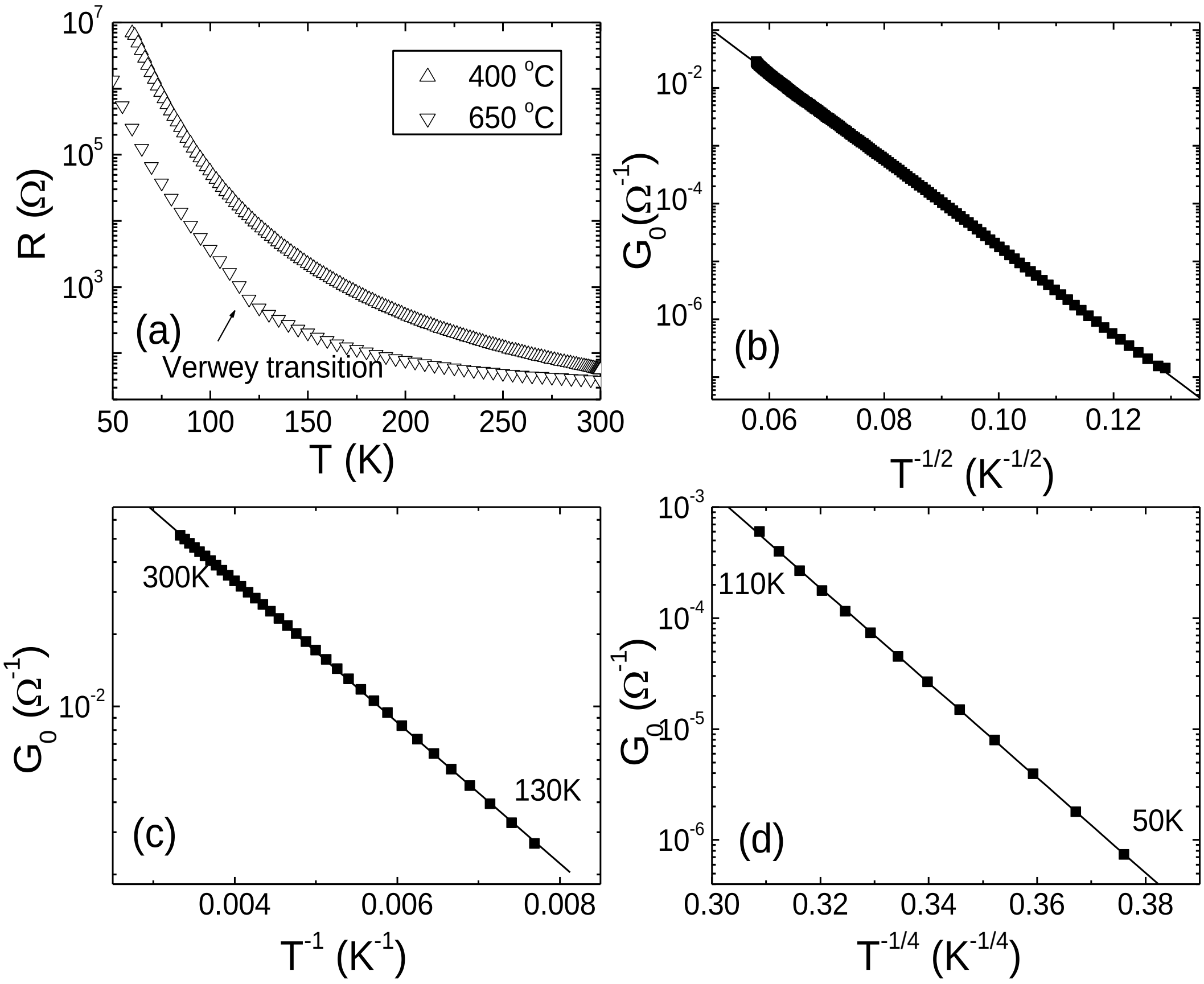

Fig. 3

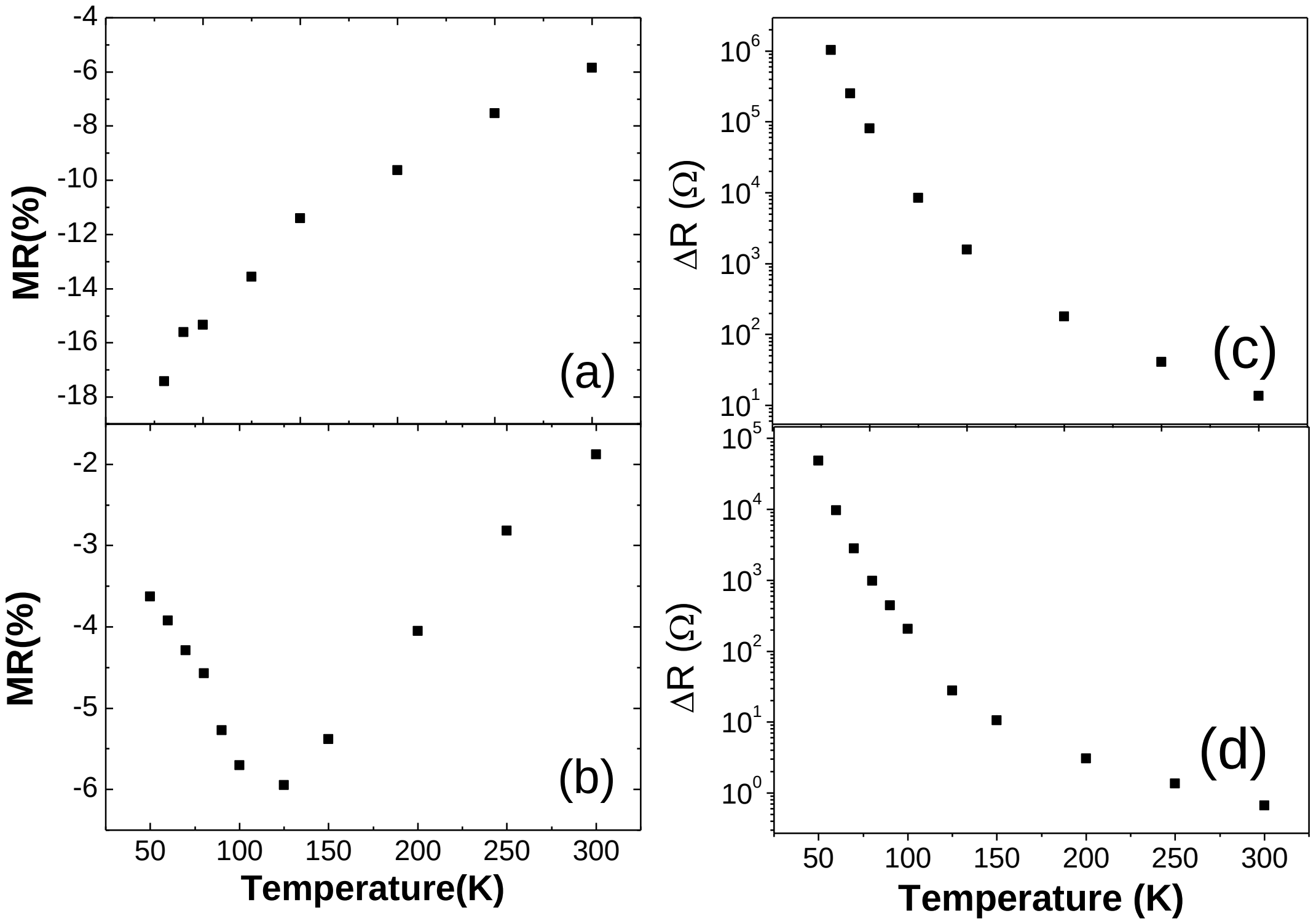

Fig. 4

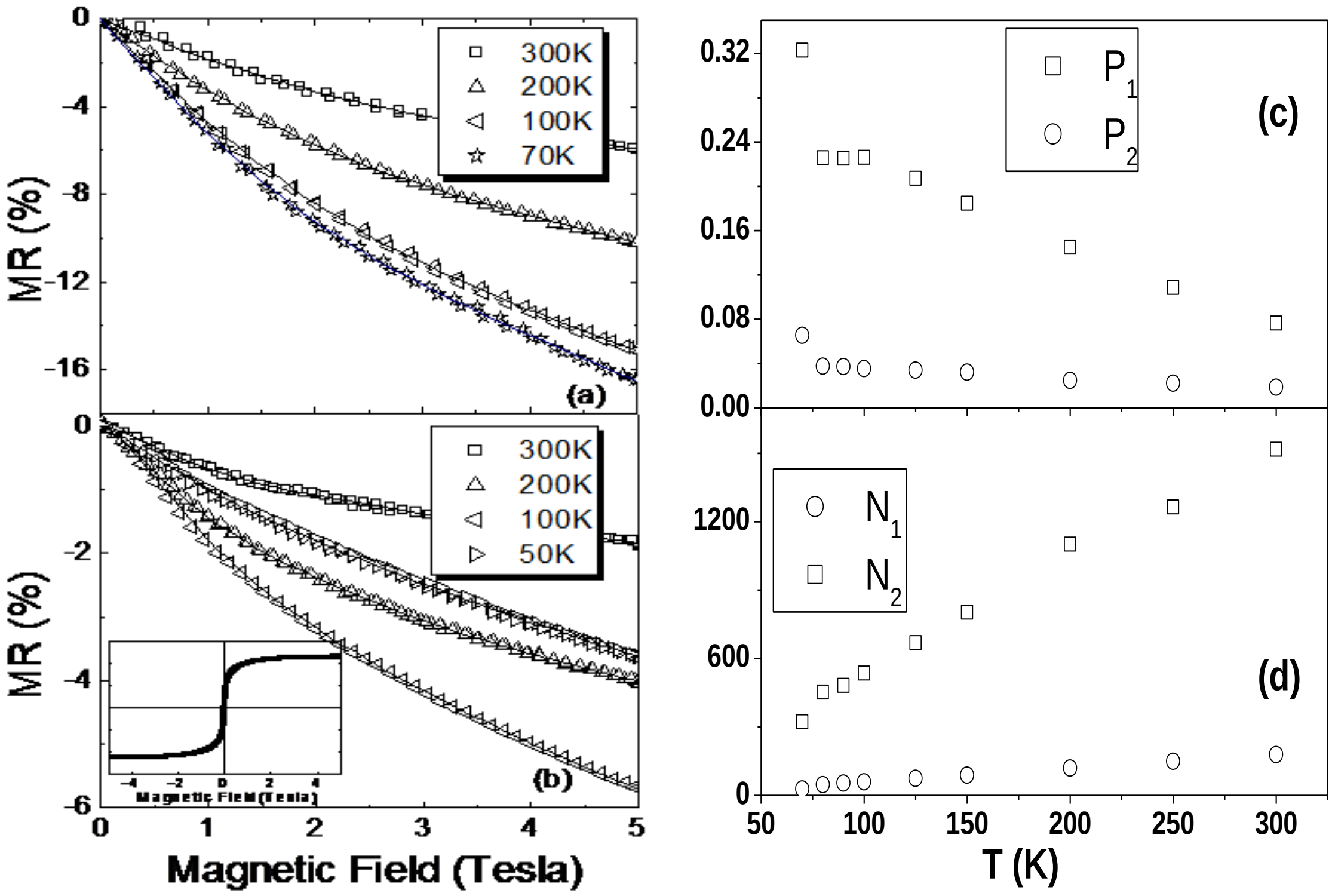

Fig. 5